\documentclass[aps,prl,showpacs]{revtex4}
\usepackage[latin1]{inputenc}
\usepackage{graphicx,amsmath,amsfonts,amssymb}
\usepackage{color}
\newcommand{\be}{\begin{equation}}
\newcommand{\ee}{\end{equation}}

\begin{document}

\title{Atomic Focusing by Quantum Fields: Entanglement Properties}
\author{I. G. da Paz\footnote{Corresponding author.}}
\affiliation{ Departamento de F\'{\i}sica, Universidade Federal do
Piau\'{\i}, Campus Ministro Petr\^{o}nio Portela, CEP 64049-550,
Teresina, PI, Brazil.}

\author{H.M. Frazão}
\affiliation{Universidade Federal do Piau\'{\i}, Campus Profa.
Cinobelina Elvas, CEP 64900-000, Bom Jesus, PI, Brazil and
Departamento de F\'{\i}sica, Instituto de Ci\^encias Exatas,
Universidade Federal de Minas Gerais, Caixa Postal 702, Belo
Horizonte, MG 30123-970, Brazil}

\author{M. C. Nemes}
\affiliation{Departamento de F\'{\i}sica, Instituto de Ci\^encias
Exatas, Universidade Federal de Minas Gerais, Caixa Postal 702, Belo
Horizonte, MG 30123-970, Brazil}

\author{J. G. Peixoto de Faria}
\affiliation{Departamento de F\'{\i}sica e Matem\'atica, Centro
Federal de Educa\c{c}\~ao Tecnol\'ogica de Minas Gerais, Av.
Amazonas 7675, Belo Horizonte, MG 30510-000, Brazil}

\begin{abstract}
The coherent manipulation of the atomic matter waves is of great
interest both in science and technology. In order to study how an
atom optic device alters the coherence of an atomic beam, we
consider the quantum lens proposed by Averbukh et al [1] to show the
discrete nature of the electromagnetic field. We extend the analysis
of this quantum lens to the study of another essentially quantum
property present in the focusing process, i.e., the atom-field
entanglement, and show how the initial atomic coherence and purity
are affected by the entanglement. The dynamics of this process is
obtained in closed form. We calculate the beam quality factor and
the trace of the square of the reduced density matrix as a function
of the average photon number in order to analyze the coherence and
purity of the atomic beam during the focusing process.

\end{abstract}

\pacs{03.75.-b, 03.65.Vf, 03.75.Be \\ \\
{\it Keywords}: Quantum lens, Atom-field entanglement, Classical
limit}

\maketitle

\section{Introduction}
Since the seminal proposal for laser cooling of atoms in dilute
gases and atom trapping \cite{Hansch}, the manipulation of all
atomic motional degrees of freedom based on the atom interaction
with external light fields have reached enormous success. Given the
recent advances in the manipulation of atoms we now observe a fast
evolution of the field both in terms of scientific knowledge and
technological applications, like in precision sensors, precise
metrology and clocks, lithography, single atom manipulation, trace
gas analysis and ultracold chemistry \cite{Schleich2010}. In
addition, the area of quantum information processing has benefited
from such advances due to the establishment of precise quantum
protocols. From the theoretical viewpoint the modeling of strongly
correlated materials and nonequilibrium quantum dynamics are
stimulating areas of research.

The dynamics of atomic beams share an intimately close analogy with
classical laser light in the paraxial approximation. The Gouy phase
discovered and measured in 1890 in the latter context is found in
any beam subject to confinement which adds a well defined phase
shift and has implications and applications in many optical systems
\cite{gouy}. The existence of a particle wave analogy to this
phenomenon has been first pointed out in Refs. \cite{Paz1} followed
by an experimental proposal in Cavity Quantum Electrodynamics (CQED)
\cite{Paz2}. Very recently this proposal has stimulated the search
for the matter wave Gouy phase in different systems: Bose-Einstein
condensates \cite{cond}, electron vortex beams \cite{elec2}, and
astigmatic electron matter waves using in-line holography
\cite{elec1}. The Gouy phase carries intrinsic properties of the
initial state and dictates the time scale of the process.

In the present work we explore the quantum version of experimental
set up proposed in Ref. \cite{Paz2} in order to show how it may be
of use to explore other quantum features as atom-field entanglement,
analysis of atomic quantum lenses proposal in \cite{Schleich1994} to
study the discrete nature of the field. The actual measurement of
this phenomenon represents a major experimental challenge, since a
quantum tomography would be required. We show here however, that the
measurement of the covariance matrix of the center of mass atomic
wavefunction indicates the presence of entanglement. Purity loss,
although far from being an easily measurable quantity is shown to
reveal the entanglement dynamics which occurs in the focusing
process. We setup a model (within experimental reach) of a focusing
and deflection of a nonresonant atomic beam propagating through a
spatially inhomogeneous quantized electromagnetic field. The
interaction of a nonresonant atom with an electromagnetic field in
the so called dispersive approximation is proportional both to the
field intensity and the susceptibility of the atom. Therefore atoms
under the influence of such fields may suffer mechanical effects
such as deviations in their center of mass motion and deflection. In
the present case we will use this property to focus atomic beams. We
address the question as to the manifestation of quantum effects in
the focusing process. In Ref. \cite{Schleich1994} the discrete
character of the photons was shown to be observable in such
experiments. Our aim within a similar scheme is to enlighten another
quantum aspect, entanglement. Interaction is the key ingredient to
produce entanglement which is an important characteristic of quantum
information protocols. We study its behavior in the atomic focusing
process.

In section II we present the model which is essentially the same as
the one used in Refs. \cite{Schleich1994,Schleich2001} with the
difference that we calculate the probability amplitude instead of
the intensities. Our procedure enable us to determine the density
matrix of the system. In section III, we present our results, in the
covariance matrix and the atom-field entanglement properties as a
function of the average photon number $\bar{n}$ showing that one
aspect of the classical limit of the field is the suppression of
entanglement as $\bar{n}$ increases. This is also apparent in the
covariance matrix. The independence of the field's granular nature
on the number of photons, shown in Refs.
\cite{Schleich1994,Schleich2001}, occurs because in that model the
authors relax the dispersive limit condition. In the present model,
we preserve the dispersive limit and the classical limit of the
field is a consequence of the disentanglement between atom and
field, apparent in the conservation of the initial purity and
coherence of the atomic beam.

\section{The Model}\label{model}
In this section we present a model that permit us focusing an atomic
beam and find an expression for the Gouy phase of matter waves that
is a connection of this phase with the inverse square of the beam
width. We consider an atomic beam propagating through a spatially
inhomogeneous quantized electromagnetic field. The atomic beam will
suffer deflection and focusing. Different Fock states deflect the
atoms in different angles and focus them at different points. We
suppose that the atomic beam is initially in a coherent Gaussian
state and obtain the equations of motion for the parameters that
characterize the structure of the wavepacket. We see that the
equations of motion is not consistent if the atomic beam was
represented at time by the one Gaussian state without the Gouy phase
term.

The model is presented in Fig. 1 in which we use the following
\citep{Schleich1994,Schleich2001}: consider two-level atoms moving
along the $Oz$ direction and that they enter in a region where a
stationary electromagnetic field is maintained. The region is the
interval $z=-L$ until $z=0$. The atomic linear momentum in this
direction is such that the de Broglie wavelength associated is much
smaller than the wavelength of the electromagnetic field. We assume
that the atomic center of mass moves classically along direction
$Oz$ and the atomic transition of interest is detuned from the mode
of the electromagnetic field (dispersive interaction). The
Hamiltonian for this model is given by
\begin{equation}
\hat{H}_{AF}=\frac{\hat{p}_{x}^{2}}{2m}+g(\hat{x})\hat{a}^{\dag}\hat{a},
\end{equation}
where $m$ is the atom mass, $\hat{p}_{x}$ and $\hat{x}$ are the
linear momentum and position along the direction $Ox$,
$\hat{a}^{\dag}$ and $\hat{a}$ are the creation and destruction
operators of a photon of the electromagnetic mode, respectively. The
coupling between atom and field is given by the function
$g(x)=\alpha \mathcal{E}^{2}(x)$ where $\alpha$ is the atomic linear
susceptibility, $\alpha=\frac{\wp^{2}}{\hbar\Delta}$, where
$\wp^{2}$ is the square of the dipole moment and $\Delta$ is the
detuning from nearest atomic resonance. $\mathcal{E}(x)$ corresponds
to the electric field amplitude in vacuum. The effective interaction
time is $t_{L}=\frac{L}{v_{z}}$, where $v_{z}$ is the longitudinal
velocity of the atoms. For simplicity the field distribution in
$z$-direction of length $L$ is assumed to have a rectangular profile
as expressed by the Heaviside step functions $\theta(z)$. The
initial width of the atomic beam is $b_{0}$ and $b_{0}^{\prime}$
represents its width at the focus.

\begin{figure}[htp]
\centering
\includegraphics[width=8 cm]{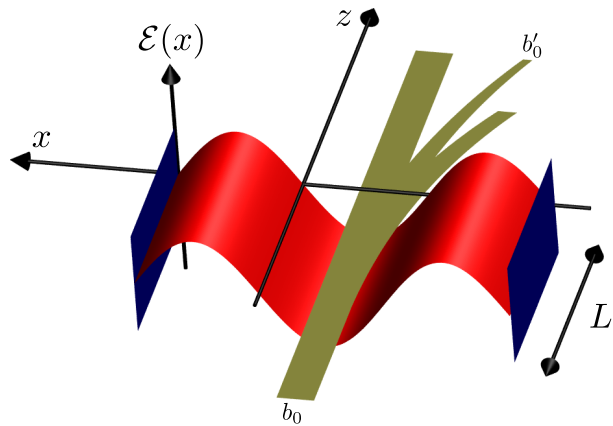}\\
\caption{Quantum lens. A beam of nonresonant atoms propagating
initially along the $z$-axis interacts with the light field in the
region $-L\leq z\leq 0$. Different Fock states deflect the atoms in
different directions and focus them at different points. The initial
width of the atomic beam is $b_{0}$ and $b_{0}^{\prime}$ represents
its width at the focus.}\label{Figure1.eps}
\end{figure}

The dynamics of the closed system is governed by the Schr\"odinger
equation
\begin{equation}
i\hbar\frac{d}{dt}|\Psi(t)\rangle=\hat{H}_{AF}|\Psi(t)\rangle.
\end{equation}
At $t=0$ the state of the system is given by a direct product of the
state corresponding to the transverse component of the atom and a
field state, $|\Psi_{CM}\rangle\otimes|\Psi_{F}\rangle$. The field
state can be expanded in the eigenstates of the number operator
$\hat{a}^{\dag}\hat{a}$
\begin{equation}
|\Psi_{F}\rangle=\sum_{n}w_{n}|n\rangle, \;\;\;
\sum_{n}|\omega_{n}|^{2}=1.
\end{equation}
When atom and field interact the atomic and field states get
entangled. We can then write
\begin{equation}
|\Psi(t)\rangle=\sum_{n}w_{n}\int_{-\infty}^{+\infty}d
x\:\psi_{n}(x,t)|x\rangle\otimes|n\rangle\:, \label{estado}
\end{equation}
where

\begin{equation}
i\hbar\frac{\partial}{\partial
t}\psi_{n}(x,t)=\left\{-\frac{\hbar^{2}}{2m}\nabla^{2}+g(x)n\right\}\psi_{n}(x,t)\:,
\label{dpsindt}
\end{equation}

or, if one defines
\begin{equation}
|\Psi_{n}(t)\rangle=\int_{-\infty}^{+\infty}d
x\;\psi_{n}(x,t)|x\rangle\:,
\end{equation}
the equation \eqref{dpsindt} takes the form
\begin{equation}
i\hbar\frac{d}{dt}|\Psi_{n}(t)\rangle=\left[\frac{\hat{p}_{x}^{2}}{2m}+g(\hat{x})n\right]|\Psi_{n}(t)\rangle.
\end{equation}
Next, we will use the harmonic approximation for $g(x)$ where we
consider that the electric field has a node in the atomic beam axis.
In addition, we considered that the width of the transverse atomic
beam $b_{0}$ is much smaller than the wavelength $\lambda$ of the
field. In this case, as a good approximation, the field creates one
square well potential for the atom in the transverse coordinate
\cite{Schleich1994,Schleich2001}. Therefore we take only the main
terms of the Taylor expansion of the function $g(x)$,
\begin{equation}
g\left(x\right) \approx g_0 - \frac{g_1^2}{2g_2} +
\frac{1}{2}g_2\left(x-x_f\right)^{2}\label{taylor} \:,
\end{equation}
where $g_{0}\equiv g(x=0)$, $g_{1}\equiv dg/dx|_{x=0}$, $g_{2}\equiv
d^{2}g/dx^{2}|_{x=0}$, $x_f\equiv -g_1/g_2$ and $\Omega_n^2 =
ng_2/m$. The combination of linear and the quadratic contributions
of the potential in a binomial reduces the problem to the motion in
the harmonic potential
$U_{n}(x)=U_{n}(x_{f})+\frac{1}{2}m\Omega_{n}^{2}(x-x_{f})^{2}$ of
the displaced harmonic oscillator with minimum
$U_{n}(x_{f})\equiv(g_{0}-g_{1}^{2}/2g_{2})n$ at $x_{f}$ and
frequency $\Omega_{n}$.

Omitting the constant term $U_{n}(x_{f})$, since it only results in
an irrelevant phase factor, we get for the Schr\"odinger equation,
\begin{eqnarray}
i\hbar\frac{d}{dt}|\Psi_{n}(t)\rangle &=&\left[\frac{\hat{p}_{x}^{2}}{2m}+\frac{1}{2}m\Omega_{n}^{2}(\hat{x}-x_{f})^{2}\right]|\Psi_{n}(t)\rangle\nonumber\\
& \equiv &\hat{H}_{n}|\Psi_{n}(t)\rangle \:. \label{Hn}
\end{eqnarray}

\subsection{Time evolution}
The general form of a Gaussian state in the position representation
is given by
\begin{equation}
   \psi \left(x,t\right)=\left(\frac{u}{\pi}\right)^{\frac{1}{4}} \exp\left(-i
    \frac{\bar{x}\bar{p}}{2\hbar}+i\mu\right)
    \exp\left[-\frac{\left(x-\bar{x}\right)^{2}\left(u + iv \right)}{2}+i\frac{\bar{p}x}{\hbar}\right],
    \label{generalgauss}
\end{equation}
where $\bar{x}$ and $\bar{p}$ are the coordinates of ``center of
mass" of the distribution in the phase space and $u$ and $v$ give
the form of this distribution. Here, $u$ is the inverse square of
the width of the Gaussian package and $v$ is related to the
curvature of the wave fronts. Different from the general form of a
Gaussian state in the position representation, defined by
Bialynicki-Birula \citep{Birula1998}, we define it in equation
\eqref{generalgauss} with an additional term $\mu$ that is a real
function of time. This global phase, in general neglected (see,
e.g., \citep{Birula1998,Piza}), has the important role of ensuring
the consistency of the equations of motion.

The dynamics governed by a Hamiltonian which is quadratic in both
position and momentum keeps the Gaussian shape of a Gaussian initial
state. This is the case of the problem treated here. The atomic
motion can be divided into two stages: the first, the atom undergoes
the action of an harmonic potential when it crosses the region of
electromagnetic field while, in the second part, the atom evolves
freely. In the two stages, the Hamiltonian governing the evolution
is quadratic in atomic position and momentum [\textit{cf.} equation
\eqref{Hn}]. Since the initial atomic state is Gaussian, we can
consider that such state will preserve the form given by equation
\eqref{generalgauss} throughout time evolution. In this case, the
parameters $\bar{x}$, $\bar{p}$, $u$, $v$ and $\mu$ are functions of
time and their respective equations of motion can be derived from
Schr\"odinger equation.

Consider a particle of mass $m$ moving under the action of an
harmonic potential. The natural frequency of this movement is
$\Omega_{n}$. The Hamiltonian governing this dynamic is given by
\begin{equation}
   \hat{H} = \frac{\hat{p}_{x}^{2}}{2m} +\frac{1}{2}m\Omega_{n}^{2} \hat{x}^{2}.
    \label{hamiltHO}
\end{equation}
In the position representation, the evolution of the state $\psi$ of
the particle is governed by the Schr\"odinger equation
\begin{equation}
    i\hbar\frac{\partial}{\partial t} \psi(x,t) =
   \left[ - \frac{\hbar^2}{2m} \frac{\partial^{2}}{\partial x^{2}} +
    \frac{1}{2}m\Omega_{n}^{2} x^{2}    \right] \psi(x,t).
    \label{Scheq}
\end{equation}
Suppose that the initial state of the particle is Gaussian. We
obtain the equations of motion for the parameters $\bar{x}$,
$\bar{p}$, $u$, $v$ and $\mu$ by substituting the general form
\eqref{generalgauss} in the equation above, grouping the terms of
same power in $\left(x-\bar{x}\right)$, and then separating the real
and imaginary parts. This procedure takes six equations for the five
parameters mentioned. The system is therefore, ``super-complete''.
Eliminating such redundancy, the equations of motion are the
following
\begin{subequations}
\label{eqmot}
  \begin{align}
    \dot{\bar{x}} &= \frac{\bar{p}}{m}\: , \label{dotx} \\
    \dot{\bar{p}} &= - m\Omega_{n}^{2} \bar{x} \:, \label{dotp} \\
   \dot{K} &= i\frac{m\Omega_{n}^{2}}{\hbar} - i\frac{\hbar}{m}
    K^{2} \: ,\label{uv}\\
    \dot{\mu}& = -\frac{\hbar}{2m}u,
  \label{dotPhi}
  \end{align}
\end{subequations}
where we define $K = u+iv$. Here, the dots indicate time derivative.
Note that the equations of motion for the coordinates of the
centroid of the distribution are equivalent to the classical
equations of movement for the position and momentum of a particle
moving in an harmonic potential. Equation \eqref{dotPhi} relates the
Gouy phase with the inverse square of the beam width. The same
result was obtained for light waves confined in the transverse
direction in Ref. \citep{feng01}. This equation does not carry any
analogy with the equation of motion for classical particles and it
is an effect of the wave behavior. If the general state
\eqref{generalgauss} does not have the parameter $\mu$, we obtain $u
= 0$. This makes no sense, since $ u $ represents the inverse square
of the width of the Gaussian package. Therefore, since the Gaussian
shape of the state has to be maintained because the dynamic is
governed by a Hamiltonian quadratic in position and momentum, the
Gouy phase term has to appear in the evolution to guarantee the
consistency of the equations of motion that represent the Gaussian
shape of the packet at a given time. The absence of the Gouy phase
term implies that the shape of the packet at a given time is a plane
wave with infinity width and not Gaussian with a finite width.

\subsection{The Focusing Process}
Here we give details of the focusing process. We consider that a
stationary electromagnetic field of wavelength $\lambda$ is produced
in an optical cavity where the relation of the wavelength of the
field and the initial width of the wavepacket in transverse
direction is such that the harmonic approximation is guaranteed. In
Fig. 2 we consider that a initial coherent Gaussian state compressed
in momentum (region I) enters in a cavity where a stationary
electromagnetic field is maintained (region II). The atoms interact
dispersively with one mode of the quantized electromagnetic field
inside the cavity. Dispersive coupling is actually one necessary
condition to produce a quantum lens, since transitions cause
aberration at the focus \citep{Schleich2001,Berman}. We note that a
key ingredient for the focusing problem is to construct inside the
cavity a compressed (squeezed) state, since the harmonic interaction
between atom and field do not produces compression and only rotates
the atomic state. When the atomic beam leaves the region of the
electromagnetic field, the atomic state evolves freely and the
compression is transferred to the position (region III).

Let us assume, as an initial atomic state, the compressed vacuum
state
\begin{equation}
\langle
x|\psi_{n}(t=0)\rangle=\psi_{n}(x,t=0)=\left(\frac{1}{b_{0}\sqrt{\pi}}\right)^{1/2}\exp\left(-\frac{x^{2}}{2b_{0}^{2}}\right)
\: ,
\end{equation}
where $b_{0}$ is the initial collimation width of the packet, which
has to be collimated in a such way to guarantee the dispersive limit
for the entire beam, i.e, $4\pi^{2} n
\Omega_{0}^{2}b_{0}^{2}/(\Delta \lambda^{2})\ll 1$, so that we can
avoid the aberration caused by the transitions and obtain a focus
with good resolution. The state above will be compressed in momentum
if $b_{0}>b_{n}=\sqrt{\hbar/\left(m\Omega_n\right)}$, where $b_{n}$
is the width of the distribution in position of the ground state of
the harmonic oscillator.

\begin{figure}[htp]
\centering
\includegraphics[width=6 cm]{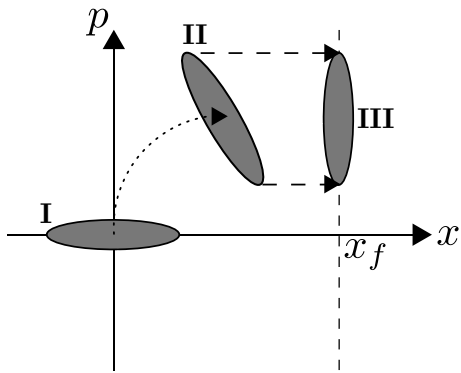}\\
\caption{Initial atomic compressed state in momentum . The evolution
inside the cavity rotates the state and transfer the compression to
the position.}\label{Figure2.eps}
\end{figure}

For the parameters $\bar{x}$, $\bar{p}$, $K$ and $\mu$, we get
\begin{equation}
\bar{x}(t<t_{L})=-x_{f}\cos\Omega_{n}t \:,
\end{equation}
\begin{equation}
\bar{p}(t<t_{L})=m\Omega_{n}x_{f}\sin\Omega_{n}t\:,
\end{equation}
and
\begin{equation}
K(t<t_{L})=\left(\cos\Omega_{n}t+i\frac{b_{n}^{2}}{b_{0}^{2}}\sin\Omega_{n}t\right)^{-1}\left(\frac{1}{b_{0}^{2}}\cos\Omega_{n}t+i\frac{1}{b_{n}^{2}}\sin\Omega_{n}t\right)\:,
\label{Kt}
\end{equation}
for the initial conditions $\bar{x}_{0}=-x_{f}, \bar{p}_{0}=0,
u=b_{0}^{-2}$ and $v=0$. Also, from equation \eqref{Kt} we obtain
\begin{equation}
u(t<t_{L})=\left[b_{0}^{2}\left(\cos^{2}\Omega_{n}t+\frac{b_{n}^{4}}{b_{0}^{4}}\sin^{2}\Omega_{n}t\right)\right]^{-1}.
\end{equation}
Now $u^{-1}$ is the width of the Gaussian wavepacket squared. At
this stage
\begin{equation}
\mu_{n}(t<t_{L})=-\frac{1}{\Omega_{n}\tau_{n}}\arctan\left[\frac{b_{n}^{2}}{b_{0}^{2}}\tan(\Omega_{n}t)\right].
\end{equation}

When the atomic beam leaves the region of the electromagnetic field,
the atomic state evolves freely. The equations of motion can be
obtained analogously and we get for $t>t_{L}$
\begin{equation}
\bar{x}(t>t_{L})=-x_{f}\cos\phi_{n}+\Omega_{n}(t-t_{L})x_{f}\sin\phi_{n},
\end{equation}
\begin{equation}
\bar{p}(t>t_{L})=m\Omega_{n}x_{f}\sin\phi_{n},
\end{equation}
\begin{equation}
K(t>t_{L})=\frac{\frac{b_{n}^{2}}{b_{0}^{2}}\cos\phi_{n}+i\sin\phi_{n}}{b_{n}^{2}\left[\cos\phi_{n}+i\frac{b_{n}^{2}}{b_{0}^{2}}\sin\phi_{n}+i\frac{t-t_{L}}{\tau_{n}}\left(\frac{b_{n}^{2}}{b_{0}^{2}}\cos\phi_{n}+i\sin\phi_{n}\right)\right]},
\end{equation}
and
\begin{equation}
b_{0}^{2}u(t>t_{L})=\left[\left(\cos\phi_{n}-\frac{t-t_{L}}{\tau_{n}}\sin\phi_{n}\right)^{2}+\frac{b_{n}^{4}}{b_{0}^{4}}\left(\sin\phi_{n}+\frac{t-t_{L}}{\tau_{n}}\cos\phi_{n}\right)^{2}\right]^{-1}\:,
\label{ut}
\end{equation}
where $\phi_{n}=\Omega_{n}t_{L}$ and $\tau_{n}=mb^{2}_{n}/\hbar$.
The focus will be located in the atomic beam region where the width
of the wavepacket is minimal. In other words, when $u(t>t_{L})$ be a
maximum there will be the focus. This will happen when the function
\begin{equation}
D(t)=\left(\cos\phi_{n}-\frac{t-t_{L}}{\tau_{n}}\sin\phi_{n}\right)^{2}+\frac{b_{n}^{4}}{b_{0}^{4}}\left(\sin\phi_{n}+\frac{t-t_{L}}{\tau_{n}}\cos\phi_{n}\right)^{2},
\label{Dt}
\end{equation}
attains its minimum value. The time for which its derivative
vanishes is given by
\begin{equation}\label{tfn}
t^{n}_{f}=\frac{z_{f}+L}{v_{z}}=t_{L}+\tau_{n}\frac{\left(1-\frac{b_{n}^{4}}{b_{0}^{4}}\right)\sin\phi_{n}\cos\phi_{n}}{\frac{b_{n}^{4}}{b_{0}^{4}}\cos^{2}\phi_{n}+\sin^{2}\phi_{n}},
\end{equation}
therefore the focus is located at
\begin{equation}
z^{n}_{f}=v_{z}\tau_{n}\frac{\left(1-\frac{b_{n}^{4}}{b_{0}^{4}}\right)\tan\phi_{n}}{\frac{b_{n}^{4}}{b_{0}^{4}}+\tan^{2}\phi_{n}}
\label{zf}.
\end{equation}
Note that different Fock state $n$ of the EM field focuses the atom
beam in different positions.

The width of the Gaussian beam that passed through the lens,
$B_{n}(t>t_{L})=1/\sqrt{u(t>t_{L})}$, can be written as
\begin{equation}
B_{n}(t>t_{L})=b^{\prime}_{0}\left[1+\left(\frac{t-t^{n}_{f}}{\tau^{\prime}_{0}}\right)^{2}\right]^{\frac{1}{2}},
\label{B}
\end{equation}
where we define
\begin{equation}
b^{\prime}_{0}=M_{n}b_{0}, \label{B0}
\end{equation}
\begin{equation}
\tau^{\prime}_{0}=M_{n}^{2}\tau_{0}, \label{tau0}
\end{equation}
and
\begin{equation}
M_{n}=\frac{1}{\sqrt{\cos^{2}\phi_{n}+\frac{b_{0}^{4}}{b_{n}^{4}}\sin^{2}\phi_{n}}}.
\label{fatorM}
\end{equation}

The prime was used here to differentiate the beam parameters after
the focusing and their parameters before the focusing. We see that
the waist of the beam is increased by factor $M_{n}$ and the time
scale $\tau_{0}$ (the analogous of the Rayleigh range for optical
beams \cite{Paz1}) is increased by the $M_{n}^{2}$. In optics, the
amount $M_{n}$ is known as magnification factor \citep{Saleh}. If
the state is not initially compressed, i.e., if $b_{n}=b_{0}$,
$M_{n}=1$ and we do not have focusing. If the state is initially
compressed in momentum, i.e., if $b_{n}<b_{0}$, $M_{n}<1$ and we
have a convergent lens. If the state is initially compressed in
position, i.e., if $b_{n}>b_{0}$, $M_{n}>1$ and we have a divergent
lens.

\section{Focusing by a Coherent State: The Generalized Uncertainty Principle and
Entanglement}\label{coerente}

In this section we will assume the field state to be in a coherent
state. Due to the interaction with the atom the atom-field
wavefunction will be an entangled state. Tracing out the field
degrees of freedom will yield a mixed density matrix for the atom.
In this case a convenient tool to describe entanglement is the
purity of this density matrix since in this case the purity loss is
directly related with the information shared between the two degrees
of freedom. Consequently the purity of this density matrix
characterize the quantum properties of the field and in practise
permit us to choose a field that focusing an atom beam and not
affect its purity. Another important parameter to define here for
the atomic beam is the analogous of the quality factor that measure
the spatial coherence of optical beam. In optics this parameter can
be defined through a covariance matrix \cite{Qiu} and for atomic
beam we will define it in the same way. The change in the initial
coherence and purity of the atomic beam manifested respectively by
the quality factor and atomic density matrix is a consequence of the
quantum nature of the field.

The density matrix corresponding to the state (\ref{estado}) is
given by \be \hat{\rho}\equiv|\psi(t)\rangle\langle\psi(t)|
=\sum_{n_{1},n_{2}} w_{n_{1}}w^{*}_{n_{2}}\int dx_{1}\int
dx_{2}\psi_{n_{1}}(x_{1},t)\psi^{*}_{n_{2}}(x_{2},t)|x_{1}\rangle\langle
x_{2}|\otimes|n_{1}\rangle\langle n_{2}|. \ee The corresponding
atomic density matrix is given by\be
\hat{\rho}_{A}\equiv\sum^{\infty}_{n=0}\langle n|\hat{\rho}|n\rangle
=\sum^{\infty}_{n=0}|w_{n}|^{2}\int dx_{1}\int dx_{2}
\psi_{n}(x_{1},t)\psi^{*}_{n}(x_{2},t)|x_{1}\rangle\langle x_{2}|.
\ee

Next we discuss the covariance matrix of the atomic beam so that we
can obtain the analogous of the beam quality factor. The covariance
matrix is defined as follows \be\label{Mc} M_C=
\left(\begin{array}{cc}
  \sigma_{xx}^{2} & \sigma_{xp} \\
  \sigma_{xp} & \sigma_{pp}^{2}
\end{array}\right),\ee where $\sigma_{xx}^{2}=\Delta x^{2}=\langle
\hat{x}^{2}\rangle-\langle \hat{x}\rangle^{2}$,
$\sigma_{pp}^{2}=\Delta p^{2}=\langle \hat{p}^{2}\rangle-\langle
\hat{p}\rangle^{2}$ are the squared variances in position and
momentum, respectively, and $\sigma_{xp}=\frac{1}{2}\langle
\hat{x}\hat{p}+\hat{p}\hat{x}\rangle-\langle \hat{x}\rangle\langle
\hat{p}\rangle$ is the position-momentum covariance. Here we obtain
for these quantities the following results
\begin{equation}
\sigma_{xx}^{2}=\sum^{\infty}_{n=0}|w_{n}|^{2}\frac{B_{n}^{2}}{2},
\end{equation}
\begin{equation}
\sigma_{pp}^{2}=\frac{\hbar^{2}}{2b_{0}^{2}}\sum_{n}\frac{|w_{n}|^{2}}{M_{n}^{2}},
\end{equation} and
\begin{equation}
\sigma_{xp}=-\frac{\hbar}{2}\sum_{n}\frac{|w_{n}|^{2}}{M_{n}^{2}}\left(\frac{t-t_{f}^{n}}{\tau_{0}}\right).
\end{equation}

The determinant of the matrix in equation (\ref{Mc}) is the
generalized Robertson-Schr\"odinger uncertainty relation and is
given by
\begin{equation}\label{schrodinger}
\sigma_{xx}^{2}\sigma_{pp}^{2}-\sigma_{xp}^{2}=\mathcal{C}\frac{\hbar^{2}}{4},
\end{equation}
where
\begin{equation}
\mathcal{C}=\sum_{n}\sum_{m}\frac{|w_{n}|^{2}|w_{m}|^{2}}{M_{m}^{2}}
\left[M_{n}^{2}+\frac{1}{M_{n}^{2}\tau_{0}^{2}}(t_{f}^{n2}-t_{f}^{n}t_{f}^{m})\right].
\label{M4}\end{equation} The constant $\mathcal{C}$ is a
proportionality constant. In wave optics it is referred to as the
squared of quality factor of the beam $\mathcal{M}^2$ \cite{fator}
and gives a measure of the spatial coherence of the laser beam. In
the matter wave context the situation is similar: when
$\mathcal{C}=1$ we will have a completely coherent and separable
atomic beam since the determinant of the covariance matrix attains
its minimum value. It may be taken as an indirect indication of
entanglement loss. This constant contains the ingredients of the
beam focusing as e.g. the ``focal time" and the field distribution.
Now let us interpret this constant and extract its physical content.
It is well known that coherent Gaussian states saturate the
generalized uncertainty principle to $\hbar^2/4$. In fact this is
true for any state subject to a quadratic dynamical evolution. We
notice that the constant $\mathcal{C}$ carriers ingredients
originated from the atom-field interaction. Therefore we expect that
for a coherent field with a sufficiently large average photon number
$\bar{n}$, entanglement with atom will become negligible and
therefore this constant should tend to one as a function of
$\bar{n}$. This is shown in Fig. 3a for the thin lens regime. The
numerical calculation was performed using parameters corresponding
to Cesium atoms in the transition $6^{2}S_{1/2}-7^{2}P_{1/2}$
\cite{Schleich1994}: wave length inside cavity
$\lambda=459\;\mathrm{nm}$, atomic mass
$m_{Cs}=2.2\times10^{-25}\;\mathrm{kg}$, Rabi frequency
$\Omega_{0}/2\pi=0.67\;\mathrm{MHz}$, cavity length
$L=100\;\mathrm{\mu m}$, longitudinal velocity
$v_{z}=300\;\mathrm{m/s}$, interaction time $t_{L}=0.3\;\mathrm{\mu
s}$ and collimation width $b_{0}/\lambda\sim1/3$. For detuning we
choose the value $\Delta=4.2\times10^8\;\mathrm{Hz}$. With these
values we obtain the following conditions
\be\phi_{n}\ll1,\;b_{n}/b_{0}\ll1\label{TL},\ee for the average
photon number $3<\bar{n}<30$. Now, neglecting in the cavity the
kinetic energy $\hat{p}_{x}^{2}/2m$ of the transverse motion of the
atom compared to the interaction energy
$g(x)\hat{a}^{\dagger}\hat{a}$ and considering the conditions above,
we obtain the regime of a thin lens \cite{Schleich1994} in which the
``focal time" is given by
\begin{equation}
t_{f}^{n}\approx t_{L}+\frac{m \Delta}{\hbar \Omega_{0}^{2}
k^{2}t_{L}n},
\end{equation}
and the magnification factor by
\begin{equation}
M_{n}\approx\frac{1}{\sqrt{1+\frac{b_{0}^{4}}{b_{n}^{4}}\phi_{n}^{2}}},
\end{equation}
where $k=2\pi/\lambda$. We see that for
small values of $\bar{n}$ ($\sim3$) where the atom field interaction
is viewed as a quantum process the values of the constant
$\mathcal{C}$ is larger than one and around $\bar{n}$ ($\sim10$)
such quantum effects are washed out.

Now a direct measure of atom-field entanglement is given by
\begin{equation}
Tr(\hat{\rho}_{A}^{2})=2\sum_{n,m}\frac{\sqrt{\tau_{0}^{\prime
m}\tau_{0}^{\prime n}}}{\sqrt{(\tau_{0}^{\prime m}+\tau_{0}^{\prime
n})^{2}+(t_{f}^{n}-t_{f}^{m})^{2}}}.
\end{equation}

This quantity, as well as $\mathcal{C}$, is time independent and
reflects the entanglement during the atom-field interaction time
$t_L$. For the same parameters as in Fig. 3a we obtain the curve in
Fig. 3b for the thin lens regime. Notice that the atomic subsystem
becomes a pure state, i.e., uncorrelated with the electromagnetic
field for sufficiently large values of $\bar{n}$ ($\ge10$).

The constant $\mathcal{C}$ in equation (\ref{M4}) may be
experimentally obtained from the quadratures of the atomic beam at
the focus so that the theoretical prediction can be tested. However
the same is not true for the purity since it depends on knowledge of
the atomic state, which may in principle, be obtained by quantum
tomography. Although this represents an enormous challenge, the
impressive progress achieved in the area in the last decade may get
us there hopefully.

\begin{figure}
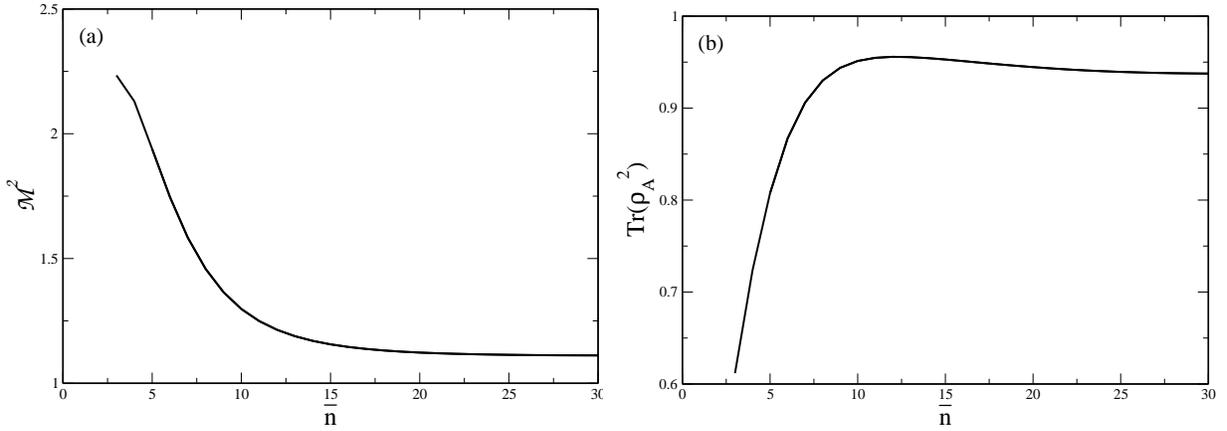

\centering
\includegraphics[width=8 cm]{Figure3a.eps}
\includegraphics[width=8 cm]{Figure3b.eps}
\caption{a) Quality factor of the atomic beam $\mathcal{M}^2$ as a
function of the average photon number $\bar{n}$ for focusing of a
Cesium atom in the thin lens regime. b) Trace of squared reduced
density matrix of atom $Tr(\hat{\rho}_{A}^2)$ as a function of the
average photon number $\bar{n}$ for focusing of a Cesium atom in the
thin lens regime. We see that its purity increases when the average
photon number increases and this is a manifestation of the
atom-field disentanglement and represents the classical limit of the
field. In a similar way, the increasing of the coherence with an
increasing average photon number in (a) can also be considered as a
manifestation of the atom-field disentanglement and consequently
represents the classical limit of the field.}\label{Figure3.eps}
\end{figure}

\section{Conclusions}

Encouraged by the recent success of the experimental measurement of
the Gouy phase for matter waves \cite{cond,elec1,elec2} proposed in
\cite{Paz2} for first time, we revisit the focusing of the atomic
beam, i.e., consider a quantum lens in order to explore other
essentially quantum features of the process. One of then has to do
with multiple foci which reflect the granular nature of the cavity
field, studied several years ago by Averbukh {\it et al} and Orzag
{\it et al} \cite{Schleich1994}. We therefore focus our attention in
the atom-field entanglement and discuss how this effect alters a
measurable quantity, i.e., the atomic beam covariance matrix and
show that a direct consequence of this entanglement is given by the
purity loss of one of the degrees of freedom. We also show that the
entanglement properties disappear by the enhancement of the average
photon number, as to be expected. In the present example, for Cesium
atoms the ``classical limit'' is reached around $\bar{n}\ge10$
depending on details of the atom-field dynamics during the
interaction time which preserve the dispersive limit such as the
detuning and the number of photon in the field state.

\begin{acknowledgments}
\section*{Acknowledgments}
We would like to thank the CNPq by financial support under grant
numbers 486920/2012-7 and 306871/2012-2. JGPF thanks support from
the program PROPESQ (CEFET/MG) under grant number PROPESQ
10122-2012.
\end{acknowledgments}



\begin{thebibliography}{99}

\bibitem{Schleich1994}

I.S. Averbukh, V.M. Akulin, W.P. Schleich, Phys. Rev. Lett. 72
(1994) 437; B. Rohwedder, M. Orszag, Phys. Rev. A 54 (1996) 5076.

\bibitem{Hansch}

T.W. H\"{a}nsch, A. Schawlow, Opt. Commun. 13 (1975) 68; D.
Wineland, H. Dehmelt, Bull. Am. Phys. Soc. 20 (1975) 637; A. Ashkin,
Phys. Rev. Lett. 40 (1978) 729; C. S. Adams, E. Riis, Prog. Quant.
Electr. 21 (1997) 1.

\bibitem{Schleich2010}

F. Schmidt-Kaler, T. Pfau, P. Schmelcher, W. Schleich, New Journal
of Phys. 12 (2010) 065014.



\bibitem{gouy}
L. G. Gouy, C. R. Acad. Sci. Paris  110 (1890) 1251; L. G. Gouy,
Ann. Chim. Phys. Ser. 6 24 (1891) 145; H. C. Kandpal, S. Raman, R.
Methrotra, Optics and Lasers in Eng. 45 (2007) 249; A. B. Ruffin, J.
V. Rudd, J. F. Whitaker, S. Feng, H. G. Winful, Phys. Rev. Lett. 83
(1999) 341; T. Feurer, N. S. Stoyanov, D. W. Ward, K. A. Nelson,
Phys. Rev. Lett. 88 (2002) 257402; F. Lindner, Phys. Rev. Lett. 92
(2004) 113001; T. Klaassen, A. Hoogeboom, M. P. van Exter, J. P.
Woerdman, J. Opt. Soc. Am. A 21 (2004) 1689; W. Zhu, A. Agrawal, A.
Nahata, Opt. Express 15 (2007) 1995; D. Kawase, Y. Miyamoto, M.
Takeda, K. Sasaki, S. Takeuchi, Phys. Rev. Lett. 101 (2008) 050501;
A. Wicht, J. M. Jensley, E. Sarajlic, S. Chu, Physica Scripta T102
(2002) 82; P. Cladé et al, Phys. Rev. A 74 (2006) 052109;

\bibitem{Paz1}

I. G. da Paz, M. C. Nemes, S. Pádua, C. H. Monken, J. G. Peixoto de
Faria, Phys. Lett. A 374 (2010) 1660; I. G. da Paz, M.C. Nemes, J.
G. Peixoto de Faria, J. Phys.: Conference Series 84 (2007) 012016.

\bibitem{Paz2}

I. G. da Paz, P. L. Saldanha, M. C.  Nemes, J. G. Peixoto de Faria,
New Journal of Phys. 13 (2011) 125005.


\bibitem{cond}

A. Hansen, J. T. Schultz, N. P. Bigelow, Conference on Coherence and
Quantum Optics Rochester, New York, United States, June 17-20, 2013.


\bibitem{elec2}

G. Guzzinati, P. Schattschneider, K. Y. Bliokh, F. Nori, Jo
Verbeeck, Phys. Rev. Lett. 110 (2013) 093601.

\bibitem{elec1}

T. C. Petersen, D. M. Paganin, M. Weyland, T. P. Simula, S. A.
Eastwood, M. J. Morgan, Phys. Rev. A 88 (2013) 043803.


\bibitem{Schleich2001}

W.P. Schleich, Quantum Optics in Phase Space, Berlin, Wiley-VCH,
2001.

\bibitem{Saleh}

B. E. A. Saleh, M. C. Teich, Fundamentals of Photonics, New York,
John Wiley et Sons, 1991.

\bibitem{Birula1998}

I. Bialynicki-Birula, Acta Phys. Polonica B. 29 (1998) 3569.

\bibitem{Piza}

A.F.R.T. Piza, Mecânica Quântica, São Paulo, Edusp, 2003.

\bibitem{feng01}

S. Fengn, H.G. Winful, Opt. Lett. 26 (2001) 485.

\bibitem{Berman}

P.R. Berman, Atom Interferometry, San Diego, Academic Press, 1997.

\bibitem{Qiu}


Y. Qiu, H. Guo, Z. Chen, Optics Communications 245 (2005) 21.

\bibitem{fator}

J.-F. Riou, W. Guerin, Y. L. Coq, M. Fauquembergue, V. Josse, P.
Bouyer, A. Aspect, Phys. Rev. Lett. 96 (2006) 070404; F. Impens,
Phys. Rev. A 77 (2008) 013619.

\end{thebibliography}
\end{document}